# Radiation condition for the 3-dimensional Helmholtz equation on the boundary of a bounded domain

Vladimir V. Gorin

vvgorin@gmail.com

**Abstract.** In the formulation of the problem of scattering of monochromatic waves and the numerical simulation of the solution to the Helmholtz equation, there is a computational inconvenience: the calculation is performed on a finite grid of discretization nodes of the finite scattering region, while the radiation conditions for the scattered wave are formulated at the infinitely distant boundary. Overcoming this inconvenience leads to a new type of boundary condition: a nonlocal boundary condition (or condition of the 4th kind).

1. **Introduction.** The main types of setting boundary value problems for equations of mathematical physics of elliptic type are well known [1-4]: these are 1) 1st boundary value problem with Dirichlet conditions – boundary condition of the 1st kind (the value of the desired function is set on the boundary of the solution domain); 2) the 2nd boundary value problem with the Neumann conditions - the boundary condition of the 2nd kind (the normal derivative of the desired function is set on the boundary); 3) The 3rd boundary value problem with Robin conditions is a boundary condition of the 3rd kind (a linear combination of the required function and its normal derivative is specified on the boundary).

All these types of conditions are linear and local (when formulating, the values of the desired function and/or its normal derivative at each point of the boundary are used separately). It should be noted that if the property of linearity of boundary conditions for linear problems is natural, then the locality of boundary conditions for fields in distributed systems is not due to anything, and is only a tradition of the historical development of problem formulations.

Non-local boundary conditions arise in a situation where a certain system of bodies is surrounded by boundless space, and it is required to solve the boundary value problem in conditional finite boundaries in such manner that the solution would not spoiled by these boundaries, and within the computational domain would coincide with the solution for boundless space. In the literature, such a problem is called the problem of constructing non-reflecting boundary conditions, and is widely studied for acoustics, hydromechanics, and scattering of waves of various nature [5–9].

The specific example of a problem with the Helmholtz equation proposed below is not new (A.A. Konstantinov and others, [8], J.B. Keller [9]). However, the solution of the problem there was done formally: in the form of a series, which turned out to be divergent and unsuitable for use in calculations. This work brings the solution to the proper form.

**2. Statement of the problem.** Consider the non-uniform Helmholtz equation in $R^3$:

$$\left(\frac{\partial^2}{\partial x^2}+\frac{\partial^2}{\partial y^2}+\frac{\partial^2}{\partial z^2}+k^2(x,y,z)\right)u=f(x,y,z). \tag{2.1}$$

Let $f(x,y,z)$ has a bounded in the $R^3$ support, which lies completely inside a closed ball $O_1$: $x^2+y^2+z^2 \leq R_1^2$ with a radius $R_1$ and a center in the origin. The parameter $k=k(x,y,z)$ has a positive constant value in the ball exterior, which is equal to $k_0$.

Thus, in the ball $O_1$ exterior, the solution $u=u(x,y,z)$ satisfies the uniform Helmholtz equation with constant coefficients. The Sommerfeld radiation conditions define the solution behavior in the infinity, namely [10]:

$$\begin{aligned} u(x,y,z) &\sim O(1/r), \quad r=\sqrt{x^2+y^2+z^2}, \\ \frac{\partial u}{\partial r}-ik_0 u &\sim o(1/r), \quad r\to\infty. \end{aligned} \tag{2.2}$$

Under conditions (2.1), (2.2) the solution of the problem in the ball $O_1$ exterior is defined *uniquely* by the boundary condition in the ball $O_1$ surface, or in the surface of a ball, which includes $O_1$ inside, because a zero boundary condition gives everywhere a zero solution in the ball $O_1$ exterior (see [11] pp. 440 – 441).

However, when setting computational boundary value problems, one has to perform calculations in a *finite* region of space, and the use of conditions (2.2) at infinity causes difficulties. So,

*Required:* to find an expression for the radiation condition *on the boundary of a limited area* (namely, here - on the boundary of the ball O: $x^2+y^2+z^2 \leq R^2$, $R > R_1$ of radius $R$, which contains the region of inhomogeneity of the original problem inside itself).

**3. Solving.** Let's move from Cartesian coordinates to spherical coordinates:

$$x=r\sin\theta\cos\varphi, \quad y=r\sin\theta\sin\varphi, \quad z=r\cos\theta. \tag{3.1}$$

Equation (2.1) is rewritten in the form[1]:

$$\left(\frac{\partial^2}{\partial r^2}+\frac{2}{r}\frac{\partial}{\partial r}+\frac{1}{r^2}\Lambda+k^2\right)u=f, \tag{3.2}$$

Here the angular part of the 3-dimensional Laplace operator is introduced

$$\Lambda=\frac{1}{\sin\theta}\frac{\partial}{\partial\theta}\sin\theta\frac{\partial}{\partial\theta}+\frac{1}{\sin^2\theta}\frac{\partial^2}{\partial\varphi^2}. \tag{3.3}$$

Outside the ball $O_1$, the solution satisfies the homogeneous equation

---
[1] For formally other functions - already from spherical coordinates - we leave the previous designations.

$$\left(\frac{\partial^2}{\partial r^2}+\frac{2}{r}\frac{\partial}{\partial r}+\frac{1}{r^2}\Lambda+k_0^2\right)u=0, \tag{3.4}$$

When the radiation conditions (2.2) are satisfied, it is represented as a series (see [4], p. 448):

$$u(r,\theta,\varphi)=\sum_{n=0}^{\infty}\sum_{m=-n}^{n}\frac{c_{nm}}{\sqrt{k_0 r}}H^{(1)}_{n+\frac{1}{2}}(k_0 r)Y_{nm}(\theta,\varphi). \tag{3.5}$$

Here $Y_{lm}(\theta,\varphi)$ are the spherical functions [12-14][2]. They are orthonormal [15]:

$$\oiint d^2 o\, Y_{lm}(\theta,\varphi)Y^*_{l'm'}(\theta,\varphi)=\delta_{ll'}\delta_{mm'}. \tag{3.6}$$

Spherical functions form a complete basis on a sphere in space $L_2(S^2)$ (see [16], [17]).

$H^{(1)}_\nu(z), H^{(2)}_\nu(z)$ are the Hankel functions of the 1-st and 2-nd kind. They have asymptotic behavior at infinity [18]:

$$H^{(1)}_\nu(z)\sim\sqrt{\frac{2}{\pi z}}\exp\left(i\left(z-\nu\frac{\pi}{2}-\frac{\pi}{4}\right)\right),\quad |z|\to\infty; \tag{3.7}$$

$$H^{(2)}_\nu(z)\sim\sqrt{\frac{2}{\pi z}}\exp\left(-i\left(z-\nu\frac{\pi}{2}-\frac{\pi}{4}\right)\right),\quad |z|\to\infty. \tag{3.8}$$

From (3.7) follows the asymptotic behavior of (3.5) at large distances

$$u(r,\theta,\varphi)\sim\sqrt{\frac{2}{\pi}}\frac{e^{ik_0 r}}{ik_0 r}\sum_{n=0}^{\infty}\sum_{m=-n}^{n}c_{nm}(-i)^n Y_{nm}(\theta,\varphi),\quad k_0 r\to\infty. \tag{3.9}$$

On the sphere $x^2+y^2+z^2=R^2$ from the obtained formulas we have:

$$u(R,\theta,\varphi)=\frac{1}{\sqrt{k_0 R}}\sum_{n=0}^{\infty}H^{(1)}_{n+\frac{1}{2}}(k_0 R)\sum_{m=-n}^{n}c_{nm}Y_{nm}(\theta,\varphi). \tag{3.10}$$

Using the formula (see [19] for the half-sum of formulas (5.6.3))

$$\frac{d}{dz}H^{(1)}_\nu(z)=H^{(1)}_{\nu-1}(z)-\frac{\nu}{z}H^{(1)}_\nu(z) \tag{3.11}$$

for the normal derivative of $u$ on the sphere, we get:

---

[2] In new sources (see Wikipedia), complex spherical functions (3.6) are denoted with subscript and superscript: $Y_l^m(\theta,\varphi)$, and two subscripts are used for real-valued spherical functions. But in this work only complex spherical functions are used, so we follow the old tradition.

$$\frac{1}{k_0}\frac{\partial u}{\partial r}(R,\theta,\varphi) = \frac{1}{\sqrt{k_0 R}} \sum_{n=0}^{\infty} \left( \frac{H^{(1)}_{n-\frac{1}{2}}(k_0 R)}{H^{(1)}_{n+\frac{1}{2}}(k_0 R)} - \frac{(n+1)}{k_0 R} \right) H^{(1)}_{n+\frac{1}{2}}(k_0 R) \sum_{m=-n}^{n} c_{nm} Y_{nm}(\theta,\varphi). \qquad (3.12)$$

Let us now construct an expression for the normal derivative of $u$ expressed in terms of the function $u$ itself on the boundary. To do this, using the orthogonality of spherical functions (3.6), we express $c_{nm}$ from (3.10) and substitute into (3.12):

$$c_{nm} = \frac{\sqrt{k_0 R}}{H^{(1)}_{n+\frac{1}{2}}(k_0 R)} \oiint d^2 o' Y^*_{nm}(\theta',\varphi') u(R,\theta',\varphi'), \qquad (3.13)$$

$$n = 0,1,\ldots \quad m = -n, -n+1, \ldots, n.$$

$$\frac{1}{k_0}\frac{\partial u}{\partial r}(R,\theta,\varphi) = \oiint d^2 o' u(R,\theta',\varphi') \sum_{n=0}^{\infty} \left( \frac{H^{(1)}_{n-\frac{1}{2}}(k_0 R)}{H^{(1)}_{n+\frac{1}{2}}(k_0 R)} - \frac{(n+1)}{k_0 R} \right) \sum_{m=-n}^{n} Y^*_{nm}(\theta',\varphi') Y_{nm}(\theta,\varphi). \qquad (3.14)$$

It can be seen from the relation obtained that the normal derivative of the desired function on the spherical boundary is expressed in terms of the values of the desired function on this boundary using the action of a linear unbounded operator on the right side of (3.14). This operator has received a name in the literature: *the Dirichlet to Neumann map* (DtN) [8]. The eigenfunctions of the operator are the spherical functions $Y_{nm}$, and the coefficients in parentheses are the spectrum of the operator. The spectrum is unbounded because the first term in parentheses in (3.14) is bounded, but the second is not. For functions that are not a finite linear combination of $Y_{nm}$, the representation (3.14) is not convenient to use and needs to be converted to a form that does not contain divergent series. Let's show how this can be done.

For spherical functions, *the addition theorem* holds [20]: if

$$\cos\gamma = \mathbf{n}\cdot\mathbf{n}' = \cos\theta\cos\theta' + \sin\theta\sin\theta'\cos(\varphi-\varphi'), \qquad (3.15)$$

is true, where

$$\mathbf{n} = \begin{bmatrix} \sin\theta\cos\varphi \\ \sin\theta\sin\varphi \\ \cos\theta \end{bmatrix}, \quad \mathbf{n}' = \begin{bmatrix} \sin\theta'\cos\varphi' \\ \sin\theta'\sin\varphi' \\ \cos\theta' \end{bmatrix}. \qquad (3.16)$$

then it holds

$$\sum_{m=-n}^{n} Y_{nm}(\theta,\varphi) Y^*_{nm}(\theta',\varphi') = \sqrt{\frac{2n+1}{4\pi}} Y_{n0}(\gamma) = \frac{2n+1}{4\pi} P_n(\cos\gamma). \qquad (3.17)$$

In what follows, for functions of angular variables $u(\theta,\varphi), u(\theta',\varphi')$, we will also use vector notation for arguments: $u(\mathbf{n}), u(\mathbf{n}')$, implying dependence on the position of a point on the surface of a sphere.

Substituting (3.17) into (3.14), we obtain

$$\frac{1}{k_0}\frac{\partial u}{\partial r}(R,\theta,\varphi) = \frac{1}{4\pi}\oiint d^2 o' u(R,\theta',\varphi') \sum_{n=0}^{\infty}\left(\frac{H^{(1)}_{n-\frac{1}{2}}(k_0 R)}{H^{(1)}_{n+\frac{1}{2}}(k_0 R)} - \frac{(n+1)}{k_0 R}\right)(2n+1)P_n(\cos\gamma). \qquad (3.18)$$

So, the desired boundary condition on a sphere of radius $R$ has the form:

$$\frac{1}{k_0}\frac{\partial u}{\partial r}(R,\mathbf{n}) = \frac{1}{4\pi}\oiint d^2 o' u(R,\mathbf{n}') DtN(k_0 R;\mathbf{n},\mathbf{n}'), \qquad (3.19)$$

where the kernel $DtN$ of the linear operator is given with expression

$$DtN(z;\mathbf{n},\mathbf{n}') = DtN(z,\xi) = \sum_{n=0}^{\infty}\left(\frac{H^{(1)}_{n-\frac{1}{2}}(z)}{H^{(1)}_{n+\frac{1}{2}}(z)} - \frac{(n+1)}{z}\right)(2n+1)P_n(\xi). \qquad (3.20)$$

The kernel $DtN$ of the operator in (3.19) is singular at $\mathbf{n}' = \mathbf{n}$, and symmetric in the angular variables $\mathbf{n}, \mathbf{n}'$. The functional series (3.20) diverges in general. This is not surprising, since the kernels of linear operators are generalized functions.

Let's divide the expression (3.20) into three components: calculation, analytical and singular:

$$DtN(z,\xi) = R(z,\xi) + A(z,\xi) + S(z,\xi), \qquad (3.21)$$

$$R(z,\xi) = i + \sum_{n=1}^{\infty}\left(\frac{H^{(1)}_{n-\frac{1}{2}}(z)}{H^{(1)}_{n+\frac{1}{2}}(z)}(2n+1) - z\left(1+\frac{1}{n}\right)\right)P_n(\xi), \qquad (3.22)$$

$$A(z,\xi) = z\sum_{n=1}^{\infty}\left(1+\frac{1}{n}\right)P_n(\xi), \qquad (3.23)$$

$$S(z,\xi) = -\frac{1}{z}\sum_{n=0}^{\infty}(n+1)(2n+1)P_n(\xi). \qquad (3.24)$$

The calculation component corresponds to a bounded operator and is determined by a series that converges absolutely and uniformly (see below). The analytical component is summed up using reference formulas (see [21] formula 6.5.1.1 at $x = \xi = \cos\gamma$, $\mu = 0$; [22] formula 8.926 A(9063.2)) and gives the integrable kernel of the bounded operator (in $L_2(S^2)$):

$$A(z,\xi) = z\left(\frac{1}{2}\sqrt{\frac{2}{1-\xi}} - 1 - \ln\sqrt{\frac{1-\xi}{2}} - \ln\left(1 + \sqrt{\frac{1-\xi}{2}}\right)\right), \tag{3.25}$$

The singular component corresponds to an unbounded integro-differential operator, the explicit form of which will be obtained below.

**4. The calculation component.** Let us prove here that the series (3.22) converges absolutely and uniformly in $\xi$, $-1 \leq \xi \leq 1$. Due to the properties of Legendre polynomials $|P_n(\xi)| \leq 1$, $-1 \leq \xi \leq 1$, it suffices to prove the convergence of the series

$$\bar{R}(z) = \sum_{n=1}^{\infty} \left| \frac{H^{(1)}_{n-\frac{1}{2}}(z)}{H^{(1)}_{n+\frac{1}{2}}(z)} (2n+1) - z\left(1 + \frac{1}{n}\right) \right| \tag{4.1}$$

for any fixed $z > 0$.

<u>Lemma.</u> For any fixed $z > 0$ there exists a finite limit

$$\lim_{\substack{n \to \infty, \\ n \in \mathbb{Z}}} n^2 \left( \frac{H^{(1)}_{n-\frac{1}{2}}(z)}{H^{(1)}_{n+\frac{1}{2}}(z)} (2n+1) - z\left(1 + \frac{1}{n}\right) \right) = c(z). \tag{4.2}$$

A corollary of the lemma is the statement about the absolute and uniform convergence of the series (3.22).

<u>Proof.</u> We use the representation of the Hankel functions of a half-integer argument in the form of series:

$$H^{(1)}_{n+\frac{1}{2}}(z) = i(-1)^{n+1} \sum_{k=0}^{\infty} \frac{(-1)^k (z/2)^{2k-\left(n+\frac{1}{2}\right)}}{\Gamma\left(k - \left(n+\frac{1}{2}\right) + 1\right) k!} + \sum_{k=0}^{\infty} \frac{(-1)^k (z/2)^{2k+n+\frac{1}{2}}}{\Gamma\left(k + n + \frac{1}{2} + 1\right) k!}. \tag{4.3}$$

It follows from the formulas [23], [24]:

$$H^{(1)}_\nu(z) \equiv \frac{J_{-\nu}(z) - e^{-\pi \nu i} J_\nu(z)}{i \sin \pi \nu}, \quad \nu \notin \mathbb{Z}, \tag{4.4}$$

$$J_\nu(x) = \sum_{k=0}^{\infty} \frac{(-1)^k}{\Gamma(k+\nu+1)\Gamma(k+1)} \left(\frac{x}{2}\right)^{2k+\nu}. \tag{4.5}$$

Here $\mathbb{Z}$ is a set of integers.

We transform expression (4.3) as follows:

$$H^{(1)}_{n+\frac{1}{2}}(z) = i(-1)^{n+1}(z/2)^{-\left(n+\frac{1}{2}\right)}\left(\sum_{k=0}^{\infty}\frac{(-1)^k(z/2)^{2k}}{\Gamma\left(k-\left(n+\frac{1}{2}\right)+1\right)k!} + i\sum_{k=0}^{\infty}\frac{(-1)^{k+n}(z/2)^{2k+2n+1}}{\Gamma\left(k+n+\frac{1}{2}+1\right)k!}\right). \quad (4.6)$$

To get away from the negative values of the argument in the gamma function in the first terms of the series, we use *the Euler complement formula* [25]:

$$\Gamma(1-\nu)\Gamma(\nu) = \frac{\pi}{\sin \pi \nu}, \quad \nu \notin \mathbb{Z}. \quad (4.7)$$

In what follows, we will also use the well-known property of the gamma function [25], called *the functional Euler equation*:

$$\Gamma(1+\nu) = \nu\Gamma(\nu). \quad (4.8)$$

We obtain

$$H^{(1)}_{n+\frac{1}{2}}(z) = -\frac{i}{\pi}(z/2)^{-\left(n+\frac{1}{2}\right)}\left(\sum_{k=0}^{\infty}\Gamma\left(n+\frac{1}{2}-k\right)\frac{(z/2)^{2k}}{k!} + i\pi\sum_{k=0}^{\infty}\frac{(-1)^k(z/2)^{2k+2n+1}}{\Gamma(k+n+3/2)k!}\right). \quad (4.9)$$

Next, we take out of brackets the largest factor $\Gamma(n+1/2)$:

$$H^{(1)}_{n+\frac{1}{2}}(z) = -\frac{i}{\pi}\Gamma(n+1/2)(z/2)^{-\left(n+\frac{1}{2}\right)}\left(\sum_{k=0}^{\infty}\frac{\Gamma(n+1/2-k)}{\Gamma(n+1/2)}\frac{(z/2)^{2k}}{k!} + \frac{i\pi}{\Gamma(n+1/2)}\sum_{k=0}^{\infty}\frac{(-1)^k(z/2)^{2k+2n+1}}{\Gamma(k+n+3/2)k!}\right). \quad (4.10)$$

The gamma function of the half-integer argument increases monotonically in absolute value with the growth of the argument everywhere, except for three values of the argument: -1/2; 1/2; 3/2, at which the absolute value of the gamma function decreases: $2\sqrt{\pi}; \sqrt{\pi}; \sqrt{\pi}/2$ respectively. For large values of $n$ in formula (4.10), this does not affect the fact that the factor from the gamma functions in the first sum in all terms of the series, except for the first, is less than 1 in absolute value. The gamma factor in the second sum has a similar property.

We select the first two terms from the first series:

$$H^{(1)}_{n+\frac{1}{2}}(z) = -\frac{i}{\pi}\Gamma(n+1/2)(z/2)^{-\left(n+\frac{1}{2}\right)} \times$$

$$\times\left(1 + \frac{(z/2)^2}{n-1/2} + \sum_{k=2}^{\infty}\frac{\Gamma(n+1/2-k)}{\Gamma(n+1/2)}\frac{(z/2)^{2k}}{k!} + \frac{i\pi(z/2)^{2n+1}}{\Gamma(n+1/2)}\sum_{k=0}^{\infty}\frac{(-1)^k(z/2)^{2k}}{\Gamma(k+n+3/2)k!}\right).$$

$$H^{(1)}_{n+\frac{1}{2}}(z) = -\frac{i}{\pi}\Gamma(n+1/2)(z/2)^{-\left(n+\frac{1}{2}\right)} \times$$

$$\times\left(1+\frac{(z/2)^2}{n-1/2}+\frac{(z/2)^4}{(n-1/2)(n-3/2)}\sum_{k=0}^{\infty}\frac{\Gamma(n-3/2-k)}{\Gamma(n-3/2)}\frac{(z/2)^{2k}}{(k+2)!}+\frac{i\pi(z/2)^{2n+1}}{\Gamma(n+1/2)}\sum_{k=0}^{\infty}\frac{(-1)^k(z/2)^{2k}}{\Gamma(k+n+3/2)k!}\right). \quad (4.11)$$

Let's denote the series:

$$r_n(z) \equiv \sum_{k=0}^{\infty}\frac{\Gamma(n-3/2-k)}{\Gamma(n-3/2)}\frac{(z/2)^{2k}}{(k+2)!}, \quad (4.12)$$

$$r'_n(z) \equiv \sum_{k=0}^{\infty}\frac{(-1)^k(z/2)^{2k}}{\Gamma(k+n+3/2)k!}. \quad (4.13)$$

These series, in any case, are bounded functions of the variable $n$ for any fixed value of $z$:

$$|r_n(z)| \leq \sum_{k=0}^{\infty}\frac{(z/2)^{2k}}{(k+2)!} = \frac{\exp\left((z/2)^2\right)-\left(1+(z/2)^2+\frac{1}{2}(z/2)^4\right)}{(z/2)^4}, \quad (4.14)$$

$$|r'_n(z)| \leq \sum_{k=0}^{\infty}\frac{(z/2)^{2k}}{k!} = \exp\left((z/2)^2\right). \quad (4.15)$$

We will also need the following estimate:

$$r_{n-1}(z) - r_n(z) = \sum_{k=0}^{\infty}\left(\frac{\Gamma(n-5/2-k)}{\Gamma(n-5/2)}-\frac{\Gamma(n-3/2-k)}{\Gamma(n-3/2)}\right)\frac{(z/2)^{2k}}{(k+2)!} =$$

$$= \sum_{k=0}^{\infty}\left(\frac{\Gamma(n-7/2-k)}{\Gamma(n-5/2)}-\frac{\Gamma(n-5/2-k)}{\Gamma(n-3/2)}\right)\frac{(z/2)^{2k+2}}{(k+3)!} =$$

$$= \sum_{k=0}^{\infty}\left(\frac{\Gamma(n-7/2)\Gamma(n-7/2-k)}{\Gamma(n-5/2)\Gamma(n-7/2)}-\frac{\Gamma(n-5/2)\Gamma(n-5/2-k)}{\Gamma(n-3/2)\Gamma(n-5/2)}\right)\frac{(z/2)^{2k+2}}{(k+3)!} =$$

$$= \sum_{k=0}^{\infty}\left(\frac{1}{n-7/2}\frac{\Gamma(n-7/2-k)}{\Gamma(n-7/2)}-\frac{1}{n-5/2}\frac{\Gamma(n-5/2-k)}{\Gamma(n-5/2)}\right)\frac{(z/2)^{2k+2}}{(k+3)!}. \quad (4.16)$$

$$|r_{n-1}(z)-r_n(z)| = \left|\sum_{k=0}^{\infty}\left(\frac{1}{n-7/2}\frac{\Gamma(n-7/2-k)}{\Gamma(n-7/2)}-\frac{1}{n-5/2}\frac{\Gamma(n-5/2-k)}{\Gamma(n-5/2)}\right)\frac{(z/2)^{2k+2}}{(k+3)!}\right| \leq$$

$$\leq \sum_{k=0}^{\infty}\left(\frac{1}{n-7/2}\left|\frac{\Gamma(n-7/2-k)}{\Gamma(n-7/2)}\right|+\frac{1}{n-5/2}\left|\frac{\Gamma(n-5/2-k)}{\Gamma(n-5/2)}\right|\right)\frac{(z/2)^{2k+2}}{(k+3)!} \leq$$

$$\leq \left( \frac{1}{n-7/2} + \frac{1}{n-5/2} \right) \frac{\exp\left((z/2)^2\right) - \left(1 + (z/2)^2 + (z/2)^4/2\right)}{(z/2)^4}. \tag{4.17}$$

It means that

$$\lim_{\substack{n\to\infty,\\ n\in\mathbb{Z}}} \left( r_{n-1}(z) - r_n(z) \right) = 0. \tag{4.18}$$

Expression (4.11) with notations (4.12), (4.13) takes a more compact form:

$$H^{(1)}_{n+\frac{1}{2}}(z) = -\frac{i}{\pi} \Gamma(n+1/2)(z/2)^{-\left(n+\frac{1}{2}\right)} f_n(z), \tag{4.19}$$

$$f_n(z) \equiv 1 + \frac{(z/2)^2}{n-1/2} + \frac{(z/2)^4}{(n-1/2)(n-3/2)} r_n(z) + \frac{i\pi(z/2)^{2n+1}}{\Gamma(n+1/2)} r'_n(z). \tag{4.20}$$

It can be seen that the function $f_n(z)$ introduced here tends to unity as $n \to \infty$.

Now, with the help of the obtained representations, let us calculate the limit (4.2).

$$c(z) = \lim_{\substack{n\to\infty,\\ n\in\mathbb{Z}}} n^2 \left( \frac{H^{(1)}_{n-\frac{1}{2}}(z)}{H^{(1)}_{n+\frac{1}{2}}(z)} (2n+1) - z\left(1 + \frac{1}{n}\right) \right) =$$

$$= \lim_{\substack{n\to\infty,\\ n\in\mathbb{Z}}} n^2 \left( \frac{-\frac{i}{\pi}\Gamma(n-1/2)(z/2)^{-\left(n-\frac{1}{2}\right)} f_{n-1}(z)}{-\frac{i}{\pi}\Gamma(n+1/2)(z/2)^{-\left(n+\frac{1}{2}\right)} f_n(z)} (2n+1) - z\left(1 + \frac{1}{n}\right) \right) =$$

$$= \lim_{\substack{n\to\infty,\\ n\in\mathbb{Z}}} n^2 \left( \frac{(z/2) f_{n-1}(z)}{(n-1/2) f_n(z)} (2n+1) - z\left(1 + \frac{1}{n}\right) \right) =$$

$$= z \lim_{\substack{n\to\infty,\\ n\in\mathbb{Z}}} n^2 \left( \frac{f_{n-1}(z)}{f_n(z)} \frac{(2n+1)}{2n-1} - \left(1 + \frac{1}{n}\right) \right) =$$

$$= z \lim_{\substack{n\to\infty,\\ n\in\mathbb{Z}}} n^2 \left( \frac{f_{n-1}(z)(2n+1)}{f_n(z)(2n-1)} - \frac{n+1}{n} \right) =$$

$$= z \lim_{\substack{n\to\infty,\\ n\in\mathbb{Z}}} n^2 \frac{f_{n-1}(z) n(2n+1) - (n+1) f_n(z)(2n-1)}{f_n(z)(2n-1)n} =$$

$$= z \lim_{\substack{n\to\infty,\\ n\in\mathbb{Z}}} n^2 \frac{f_{n-1}(z)(2n^2+n) - f_n(z)(2n^2+n-1)}{f_n(z)(2n-1)n} =$$

$$= z \lim_{\substack{n \to \infty, \\ n \in \mathbb{Z}}} n^2 \frac{(2n^2+n)(f_{n-1}(z)-f_n(z))+f_n(z)}{f_n(z)(2n-1)n} =$$

$$= \frac{z}{2} + \frac{z}{2} \lim_{\substack{n \to \infty, \\ n \in \mathbb{Z}}} (2n^2+n)(f_{n-1}(z)-f_n(z)) =$$

$$= \frac{z}{2} + \frac{z}{2} \lim_{\substack{n \to \infty, \\ n \in \mathbb{Z}}} (2n^2+n) \left( 1 + \frac{(z/2)^2}{n-3/2} + \frac{(z/2)^4}{(n-3/2)(n-5/2)} r_{n-1}(z) + \frac{i\pi(z/2)^{2n-1}}{\Gamma(n-1/2)} r'_{n-1}(z) - \right.$$

$$\left. -1 - \frac{(z/2)^2}{n-1/2} - \frac{(z/2)^4}{(n-1/2)(n-3/2)} r_n(z) - \frac{i\pi(z/2)^{2n+1}}{\Gamma(n+1/2)} r'_n(z) \right) =$$

$$= \frac{z}{2} + \frac{z}{2} \lim_{\substack{n \to \infty, \\ n \in \mathbb{Z}}} (2n^2+n) \left( \frac{(z/2)^2}{n-3/2} - \frac{(z/2)^2}{n-1/2} + \right.$$

$$+ \frac{(z/2)^4}{(n-3/2)(n-5/2)} r_{n-1}(z) - \frac{(z/2)^4}{(n-1/2)(n-3/2)} r_n(z) +$$

$$\left. + \frac{i\pi(z/2)^{2n-1}}{\Gamma(n-1/2)} r'_{n-1}(z) - \frac{i\pi(z/2)^{2n+1}}{\Gamma(n+1/2)} r'_n(z) \right).$$

The last parenthesized line in the limit disappears because the gamma function grows faster than both the power function and the exponent. Next, we get

$$c(z) = \frac{z}{2} + \left(\frac{z}{2}\right)^3 \lim_{\substack{n \to \infty, \\ n \in \mathbb{Z}}} \left( \frac{(2n^2+n)}{(n-3/2)(n-1/2)} + (z/2)^2 \left( \frac{(2n^2+n)r_{n-1}(z)}{(n-3/2)(n-5/2)} - \frac{(2n^2+n)r_n(z)}{(n-1/2)(n-3/2)} \right) \right) =$$

$$= \frac{z}{2} + \left(\frac{z}{2}\right)^3 \left( 2 + 2(z/2)^2 \lim_{\substack{n \to \infty, \\ n \in \mathbb{Z}}} (r_{n-1}(z) - r_n(z)) \right) = \frac{z}{2} + 2 \left(\frac{z}{2}\right)^3.$$

In this way,

$$\lim_{\substack{n \to \infty, \\ n \in \mathbb{Z}}} n^2 \left( \frac{H^{(1)}_{n-\frac{1}{2}}(z)}{H^{(1)}_{n+\frac{1}{2}}(z)} (2n+1) - z \left(1 + \frac{1}{n}\right) \right) = \frac{z}{2} + 2 \left(\frac{z}{2}\right)^3. \tag{4.21}$$

The lemma is proven.

As already mentioned, this proves the absolute and uniform convergence of the series (3.22), which determines the computational part of the Dirichlet-Neumann operator.

**5. The singular part of the operator DtN.** Expression (3.24) admits the following further processing. To obtain a convergent series, one should use the formula [14] for one of the properties of spherical functions:

$$\Lambda' Y^*_{nm}(\theta', \varphi') = -n(n+1) Y^*_{nm}(\theta', \varphi'). \tag{5.1}$$

Here $\Lambda'$ means the operator (3.3) taken with respect to the variables $\theta', \varphi'$. Using formula (3.17), we represent expression (3.24) as

$$S(z, \xi) = S(z; \mathbf{n}, \mathbf{n}') = -\frac{4\pi}{z} \sum_{n=0}^{\infty} (n+1) \sum_{m=-n}^{n} Y_{nm}(\mathbf{n}) Y^*_{nm}(\mathbf{n}'). \tag{5.2}$$

This is the spectral expansion of the singular part S of the operator DtN in terms of its eigenfunctions, the spherical functions[3].

Using (5.1), we substitute into (5.2) the expression

$$Y^*_{nm}(\theta', \varphi') = \Lambda' \frac{-1}{n(n+1)} Y^*_{nm}(\theta', \varphi'), \quad n = 1, 2, \ldots \tag{5.3}$$

For transformations, we use the well-known formulas (see [22] formula 6.5.1.1 at $x = \cos\gamma$, $\mu = 0$; also [23] formula 8.926 A(9063.2)):

$$\sum_{n=0}^{\infty} P_n(\cos\gamma) = \frac{1}{2}\csc\frac{\gamma}{2}, \quad P_0(\cos\gamma) = 1, \quad \sum_{n=1}^{\infty} \frac{1}{n} P_n(\cos\gamma) = -\ln\sin\frac{\gamma}{2} - \ln\left(1 + \sin\frac{\gamma}{2}\right), \tag{5.4}$$

$$\sin\frac{\gamma}{2} = \sqrt{\frac{1-\cos\gamma}{2}} = \sqrt{\frac{1-\mathbf{n}\cdot\mathbf{n}'}{2}}. \tag{5.5}$$

We obtain

$$S(z; \mathbf{n}, \mathbf{n}') = -\frac{1}{z} + \frac{1}{z}\Lambda' F(\mathbf{n} \cdot \mathbf{n}'), \tag{5.6}$$

$$F = \frac{1}{\sin\frac{\gamma}{2}} - \ln\sin\frac{\gamma}{2} - \ln\left(1+\sin\frac{\gamma}{2}\right) = \sqrt{\frac{2}{1-\xi}} - \ln\sqrt{\frac{1-\xi}{2}} - \ln\left(1+\sqrt{\frac{1-\xi}{2}}\right). \tag{5.7}$$

Let us note here that in representation (5.6) the kernel *S* formally ceased to be a symmetric function of the scalar product of vectors $\mathbf{n}, \mathbf{n}'$, since the angular operator $\Lambda'$ acts on the variables $\theta', \varphi'$, but does not act on the variables $\theta, \varphi$. Symmetry had to be sacrificed temporarily in order to get away from the divergent expression (3.24). The symmetry will be restored farther, and the way is rather interesting.

The angular part of the Laplace operator (3.3) can be expanded into a product of first-order differential operators acting on the sphere:

---
[3] The expansion of the operator DtN itself is given by formula (3.14).

$$\Lambda = \operatorname{div} \nabla, \tag{5.8}$$

Here $\nabla$ means the gradient operator on the sphere

$$\nabla u(\theta, \varphi) \equiv \left( \frac{\partial u}{\partial \theta} \quad \frac{1}{\sin \theta} \frac{\partial u}{\partial \varphi} \right), \tag{5.9}$$

The divergence operator on the sphere has the form

$$\operatorname{div} \mathbf{v} \equiv \frac{1}{\sin \theta} \frac{\partial}{\partial \theta} (\sin \theta \, v_\theta) + \frac{1}{\sin \theta} \frac{\partial}{\partial \varphi} v_\varphi, \tag{5.10}$$

here $\mathbf{v}$ is an arbitrary vector field on a sphere of unit radius

$$\mathbf{v} = \left( v_\theta(\theta, \varphi) \quad v_\varphi(\theta, \varphi) \right), \quad \mathbf{v}^2 = v_\theta^{\,2} + v_\varphi^{\,2}. \tag{5.11}$$

Note that the sphere has no boundary, and all integrals of the total divergence of smooth vector functions vanish. Operator kernels have singularities; when integrating them, all expressions should be understood as a *weak limit of regularized expressions* [4]. Representation (5.8) makes it possible, by integrating twice by parts, to permute the operators:

$$v(\mathbf{n}) = \oiint d^2 o' \left( \Lambda' F(\mathbf{n} \cdot \mathbf{n}') \right) u(\mathbf{n}') = \oiint d^2 o' F(\mathbf{n} \cdot \mathbf{n}') \Lambda' u(\mathbf{n}'). \tag{5.12}$$

Therefore, in view of representation (5.12), expression (5.6) can also be written in the form:

$$S(z; \mathbf{n}, \mathbf{n}') = -\frac{1}{z} + \frac{1}{z} F(\mathbf{n} \cdot \mathbf{n}') \Lambda', \tag{5.13}$$

where it is understood that the differential operator $\Lambda'$ does not act on $F$, but on a sufficiently smooth function that will stand on the right. The usefulness of expression (5.13) is that $F(\mathbf{n} \cdot \mathbf{n}')$ contains a singularity already *integrable* on the sphere at $\mathbf{n}' = \mathbf{n}$, while the integral of (5.6) diverges.
Expression (5.13) shows that the operator

$$S = \frac{1}{z}(F\Lambda - I), \tag{5.14}$$

where

$$Fu(\mathbf{n}) = \oiint d^2 o' F(\mathbf{n} \cdot \mathbf{n}') u(\mathbf{n}'), \tag{5.15}$$

$$Iu(\mathbf{n}) = \oiint d^2 o' u(\mathbf{n}'), \tag{5.16}$$

is an integro-differential operator of the 2nd (differential) order. All three operators: $\hat{F}, \Lambda, I$, included in the right side of the expression (5.14), have the same set of eigenfunctions, these are spherical functions $Y_{nm}(\theta, \varphi)$. Therefore, the operator factors in (5.14) commute with each other and are self-adjoint operators, just like the operator (5.14) itself. The boundary condition with the operator of the 2nd differential order along the boundary, however, is not convenient for applying the *finite element method* (FEM) in calculations, since the idea of the method is to apply the so-called *semi-weak* form of a boundary value problem for second-order partial differential equations (see J. Descloux [26], also [27]). It allows to bypass the calculation of the second derivatives of the desired function in a process of numerical solving. Therefore, it is highly desirable to find a representation for a new type of boundary conditions also in a *semi-weak* form, that is, not containing second derivatives of the desired function along the boundary.

---

[4] The regularization here is performed in such a way that at $\xi = \pm 1$, the kernel has a finite value, and the derivative vanishes.

**6. Semi-weak form of the boundary conditions.** Consider the bilinear form[5]:

$$B(u,v) = \oiint d^2 o\, v^*(\mathbf{n}) \oiint d^2 o'\, F(\mathbf{n}\cdot\mathbf{n}') \Lambda' u(\mathbf{n}') = (\hat{F}\Lambda u, v). \tag{6.1}$$

Is it possible to convert this form to a symmetrical view

$$B_1(u,v) = \oiint d^2 o \oiint d^2 o'\, \frac{\partial v^*}{\partial \tau}(\mathbf{n},\mathbf{n}') b(\mathbf{n}\cdot\mathbf{n}') \frac{\partial u}{\partial \tau'}(\mathbf{n},\mathbf{n}') \equiv \left(\hat{b}\frac{\partial u}{\partial \tau}, \frac{\partial v}{\partial \tau}\right), \tag{6.2}$$

including only the first derivatives of functions $u$, $v$? Here $\boldsymbol{\tau}, \boldsymbol{\tau}'$ are the vectors tangent to the surface of the unit sphere at the points $\mathbf{n}, \mathbf{n}'$ and directed towards each other tangentially along the arc of the great circle connecting these points. Directional derivatives are defined along them

$$\frac{\partial v^*}{\partial \tau} = \frac{\partial v^*}{\partial \theta}(\theta,\varphi)\cos\alpha + \frac{1}{\sin\theta}\frac{\partial v^*}{\partial \varphi}(\theta,\varphi)\sin\alpha, \tag{6.3}$$

$$\frac{\partial u}{\partial \tau'} = \frac{\partial u}{\partial \theta'}(\theta',\varphi')\cos\alpha' + \frac{1}{\sin\theta'}\frac{\partial u}{\partial \varphi'}(\theta',\varphi')\sin\alpha'. \tag{6.4}$$

Angles $\alpha, \alpha'$ are the rotation angles of the vectors $\boldsymbol{\tau}, \boldsymbol{\tau}'$ relative to the meridians $\varphi = \text{const}$ and $\varphi' = \text{const}$ of the selected coordinate system on the sphere, respectively. It can be shown that they are determined by the relations

$$\cos\alpha = \frac{1}{\sqrt{1-\xi^2}}\frac{\partial \xi}{\partial \theta} = -\frac{\partial \arccos\xi}{\partial \theta} = -\frac{\partial \gamma}{\partial \theta}, \tag{6.5}$$

$$\sin\alpha = \frac{1}{\sin\theta}\frac{1}{\sqrt{1-\xi^2}}\frac{\partial \xi}{\partial \varphi} = -\frac{1}{\sin\theta}\frac{\partial \gamma}{\partial \varphi}; \tag{6.6}$$

$$\cos\alpha' = -\frac{\partial \gamma}{\partial \theta'}, \tag{6.7}$$

$$\sin\alpha' = -\frac{1}{\sin\theta'}\frac{\partial \gamma}{\partial \varphi'}. \tag{6.8}$$

---

[5] Here the last expression implies the scalar product in the Hilbert space $L_2(S^2)$ of functions integrable on the sphere of unit radius with modulus squared.

Here the function $\gamma = \gamma(\theta, \varphi; \theta', \varphi')$ is defined from (3.15). For directional derivatives (6.3) and (6.4) we obtain

$$\frac{\partial v^*}{\partial \tau}(\mathbf{n}, \mathbf{n}') = -\left(\frac{\partial v^*}{\partial \theta}\frac{\partial \gamma}{\partial \theta} + \frac{1}{\sin^2 \theta}\frac{\partial v^*}{\partial \varphi}\frac{\partial \gamma}{\partial \varphi}\right) = -\nabla v^* \cdot \nabla \gamma, \tag{6.9}$$

$$\frac{\partial u}{\partial \tau'}(\mathbf{n}, \mathbf{n}') = -\left(\frac{\partial u}{\partial \theta'}\frac{\partial \gamma}{\partial \theta'} + \frac{1}{\sin^2 \theta'}\frac{\partial u}{\partial \varphi'}\frac{\partial \gamma}{\partial \varphi'}\right) = -\nabla' u \cdot \nabla' \gamma. \tag{6.10}$$

From relations $\xi = \cos\gamma$ and (3.15) one can obtain useful relations:

$$(\nabla \xi)^2 = (1 - \xi^2), \tag{6.11}$$

$$(\nabla \gamma)^2 = 1. \tag{6.12}$$

$$\Lambda \gamma = \frac{\xi}{\sqrt{1-\xi^2}} = \cot\gamma. \tag{6.13}$$

Let us substitute representation (6.9) into expression (6.2) and integrate by parts in the first integral

$$B_1(u,v) = \oiint d^2 o\, v^* \oiint d^2 o'\, \frac{\partial u}{\partial \tau'} \sqrt{1-\xi^2}\left(b\frac{\xi}{1-\xi^2} - b'\right) - \oiint d^2 o\, \mathrm{div}\left(v^* \oiint d^2 o'(\nabla\gamma) b \frac{\partial u}{\partial \tau'}\right).$$

$$B_1(u,v) = -\oiint d^2 o\, v^* \oiint d^2 o'\, \frac{\partial u}{\partial \tau'}\frac{d}{d\xi}\left(b(\xi)\sqrt{1-\xi^2}\right) - \oiint d^2 o\, \mathrm{div}\left(v^* \oiint d^2 o'(\nabla\gamma) b \frac{\partial u}{\partial \tau'}\right). \tag{6.14}$$

But, on the other hand, from definition (6.1), the same form is expressed in terms of the function $F$ obtained earlier:

$$B(u,v) = \oiint d^2 o\, v^* \oiint d^2 o'\, \mathrm{div}'(F(\xi)\nabla' u(\mathbf{n}')) - \oiint d^2 o\, v^* \oiint d^2 o'\, F'(\xi)\sqrt{1-\xi^2}\frac{\partial u}{\partial \tau'}. \tag{6.15}$$

Comparison of expressions (6.14) and (6.15), provided that terms with total divergence under the integral vanish[6], and taking into account arbitrary functions $u$, $v$, leads to the equation

$$b'(\xi) - b(\xi)\frac{\xi}{1-\xi^2} = F'(\xi),$$

$$\frac{d}{d\xi}\left(b(\xi)\sqrt{1-\xi^2}\right) = F'(\xi)\sqrt{1-\xi^2}, \tag{6.16}$$

which defines the desired function $b$. The general solution of this ODE has the form

---

[6] The vanishing depends on the nature of the singularity of the expressions for $b$ and $F$ at $\xi \to +1$, since the singular point on the sphere creates the boundary of the integration region, which is absent for smooth functions.

$$b(\xi)\sqrt{1-\xi^2} = C + \int_{-1}^{\xi} d\xi' F'(\xi')\sqrt{1-\xi'^2}. \tag{6.17}$$

Substituting the expression (5.7) for *F* and performing the integration gives the expression

$$b(\xi) = \frac{2\operatorname{arcth}\sqrt{\frac{1+\xi}{2}}}{\sqrt{1-\xi^2}} - 1 + \frac{C}{\sqrt{1-\xi^2}}. \tag{6.18}$$

$$b(\cos\gamma)\sin\gamma = 2\operatorname{arcth}\cos\frac{\gamma}{2} - \sin\gamma, \quad C = 0. \tag{6.19}$$

If we put $C = 0$, then the expression for *b* vanishes at $\xi = -1$, and the expression has only one (integrable) singularity at $\xi = +1$. For $C \neq 0$, this expression has two (integrable) singularities at $\xi = \pm 1$. The requirement that there be no singularity at $\xi = -1$ is quite natural, since the function *F* does not have a singularity also at this antipode point.

Fig. 1 shows how expression (6.19) can be *regularized* and should be understood as a weak regularization limit.

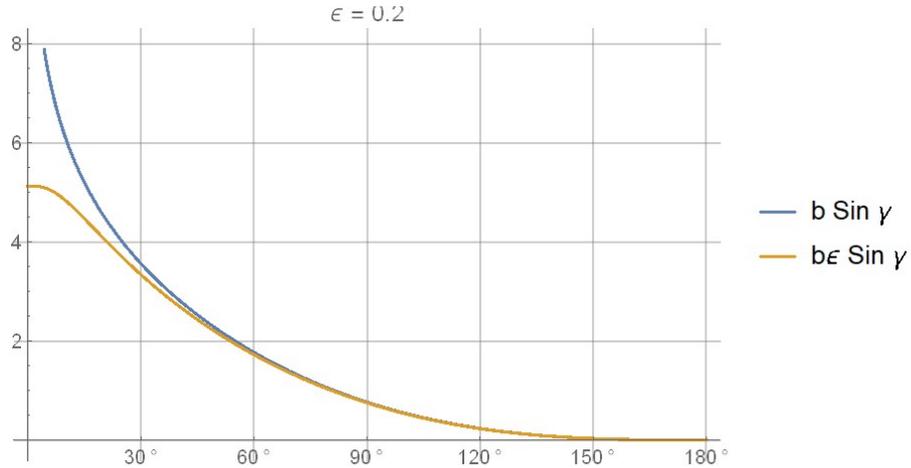

**Fig. 1.** Regularization effect for *b*.

Studies have shown that terms with total divergence in expressions (6.14), (6.15) vanish in the weak limit.

**7. Sommerfeld asymptotic for the operator DtN.** At large distances $z \gg 1$, the role of the singular part S (5.14) or (3.24) rapidly decreases with increasing *z* due to the factor $1/z$. Let us analyze the properties of the regular part: the sum $K \equiv R + A$ of expressions (3.22) and (3.23)

$$K(z,\xi) = \sum_{n=0}^{\infty} \frac{H^{(1)}_{n-\frac{1}{2}}(z)}{H^{(1)}_{n+\frac{1}{2}}(z)}(2n+1)P_n(\xi). \tag{7.1}$$

The regular part has the property of localization of the kernel as the dimensionless radius $z$ of the boundary sphere increases. Fig. 2 shows the behavior of the regular part of the mapping DtN: as $z$ increases, it becomes more and more localized to the delta function (see below).

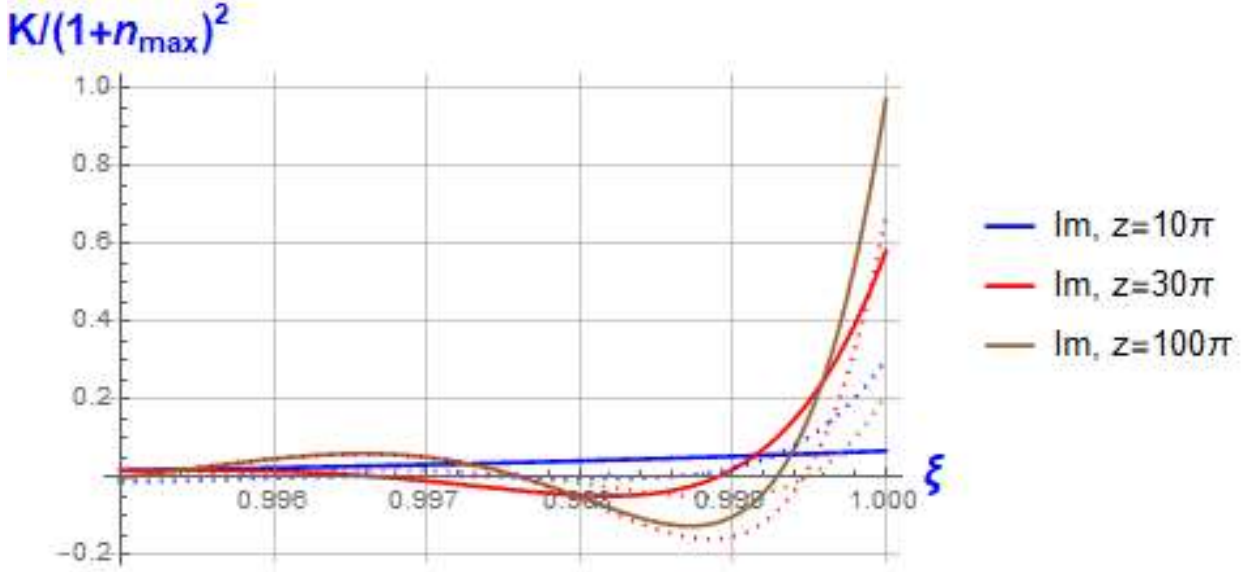

**Fig. 2.** Numerical calculations of the imaginary (solid line) and real (dotted line) parts of the regular part K of the operator DtN divided by $\operatorname{Im} K(\infty, 1; n_{max}) = \sum_{n=0}^{n_{max}} (2n+1) = (n_{max}+1)^2$. The first 402 terms of series (3.22) at $z = 10\pi$ (blue), $z = 30\pi$ (red), and $z = 100\pi$ (brown) are included.

The Sommerfeld conditions (2.2) imply not only a large (dimensionless) distance $z = k_0 R$, $z \gg 1$ to the vicinity of the observation point, but also a fairly small size of the region $z_1 = k_0 R_1$, $z_1 \ll z$ in which the radiation source is concentrated. Indeed, if we imagine that $z_1$ also increases infinitely together with $z$, then it becomes obvious that the second of conditions (2.2) may not be satisfied. For example, two point sources of a field and an observation point form an equilateral triangle, the size of which (in wavelengths) increases indefinitely. At the observation point, we will get a picture of the interference of two plane waves propagating at an angle of 60 degrees to each other, and such a field does not correspond to (2.2). Also, the first of the conditions (2.2) may not be satisfied if the field sources are allowed to be at a finite distance from the boundary sphere as its radius increases (that is, seek the limit $z \to \infty$ at $z - z_1 = \text{const}$). Obviously, conditions (2.2) must be satisfied when passing to the limit $z \to \infty$, $z_1 = \text{const}$. But this requirement is stronger than the one under which the problem was solved: $0 < R_1 < R$ (or in dimensionless quantities $0 < z_1 < z$). The effect of an *additional* requirement leading to conditions (2.2) is *to cut off* the high-order angular modes[7] *before* passing to the limit $z \to \infty$, which masks the effect of the lack of uniform convergence of the series $K$ (7.1) defining the operator kernel. This additional requirement makes it possible to replace the Hankel functions in the representation (7.1) at large distances by their well-known asymptotic expression (3.7), which substitutes (7.1) with its asymptotic representation

$$K(\infty,\xi) = i\sum_{n=0}^{\infty}(2n+1)P_n(\xi) = 2i\delta(1-\xi). \tag{7.2}$$

i.e., leads the kernel to localization in angles, and (3.19) to the formula for a plane wave. Indeed, we use the formulas:

$$\sum_{n=0}^{\infty}\sum_{m=-n}^{n} Y_{nm}(\mathbf{n})Y_{nm}^*(\mathbf{n}') = \frac{1}{4\pi}\sum_{n=0}^{\infty}(2n+1)P_n(\mathbf{n}\cdot\mathbf{n}'),$$

$$\frac{1}{2}\sum_{n=0}^{\infty}(2n+1)P_n(\xi) = \delta(1-\xi). \tag{7.3}$$

$$\sum_{n=0}^{\infty}\sum_{m=-n}^{n} Y_{nm}(\mathbf{n})Y_{nm}^*(\mathbf{n}') = \delta^2(\mathbf{n},\mathbf{n}'). \tag{7.4}$$

---

[7] The number of significant angular modes is proportional to the number of wavelengths that fit on the length of the great circle of the ball $O_1$.

The first equality is the addition theorem (3.17). Equality (7.3) is the series expansion of the delta function in Legendre polynomials, it follows from the orthogonality and normalization of Legendre polynomials [13-15]:

$$\int_{-1}^{1} d\xi P_m(\xi)P_n(\xi) = \delta_{mn}\frac{2}{2n+1}.$$

Indeed, for any continuous function $f(\xi)$, $-1 \le \xi \le 1$ we have:

$$\int_{-1}^{1} d\xi f(\xi)\frac{1}{2}\sum_{n=0}^{\infty}(2n+1)P_n(\xi) = \frac{1}{2}\sum_{n=0}^{\infty}(2n+1)\int_{-1}^{1} d\xi f(\xi)P_n(\xi) =$$

$$= \frac{1}{2}\sum_{n=0}^{\infty}(2n+1)\int_{-1}^{1} d\xi\left(\sum_{m=0}^{\infty} f_m P_m(\xi)\right)P_n(\xi) =$$

$$= \frac{1}{2}\sum_{n=0}^{\infty}(2n+1)\sum_{m=0}^{\infty} f_m \int_{-1}^{1} d\xi P_m(\xi)P_n(\xi) = \sum_{n=0}^{\infty}\sum_{m=0}^{\infty} f_m\delta_{nm} = \sum_{n=0}^{\infty} f_n = \sum_{n=0}^{\infty} f_n P_n(1) = f(1).$$

But by definition we have also $\int_{-1}^{1} d\xi f(\xi)\delta(1-\xi) = f(1)$.

Equality (7.4) is the expansion of a two-dimensional delta function on a sphere into a series in terms of spherical functions; it follows from the orthogonality and normalization (3.6) of spherical functions. Indeed, we rewrite (3.6) as

$$\oiint d^2o'\, Y_{lm}(\mathbf{n}')Y^*_{l'm'}(\mathbf{n}') = \delta_{ll'}\delta_{mm'}.$$

Multiply by $Y_{l'm'}(\mathbf{n})$ and sum over primed indices

$$\sum_{l'=0}^{\infty}\sum_{m'=-l'}^{l'} Y_{l'm'}(\mathbf{n}) \oiint d^2o' Y_{lm}(\mathbf{n}') Y^*_{l'm'}(\mathbf{n}') = \sum_{l'=0}^{\infty}\sum_{m'=-l'}^{l'} Y_{l'm'}(\mathbf{n})\delta_{ll'}\delta_{mm'} = Y_{lm}(\mathbf{n}).$$

Let's represent the result in the form

$$Y_{lm}(\mathbf{n}) = \oiint d^2o' Y_{lm}(\mathbf{n}') \sum_{l'=0}^{\infty}\sum_{m'=-l'}^{l'} Y_{l'm'}(\mathbf{n}) Y^*_{l'm'}(\mathbf{n}') = \oiint d^2o' X(\mathbf{n},\mathbf{n}') Y_{lm}(\mathbf{n}'),$$

$$X(\mathbf{n},\mathbf{n}') \equiv \sum_{l'=0}^{\infty}\sum_{m'=-l'}^{l'} Y_{l'm'}(\mathbf{n}) Y^*_{l'm'}(\mathbf{n}').$$

Multiply by $f_{lm}$ and sum over all indices

$$f(\mathbf{n}) = \sum_{l=0}^{\infty}\sum_{m=-l}^{l} f_{lm} Y_{lm}(\mathbf{n}) = \sum_{l=0}^{\infty}\sum_{m=-l}^{l} f_{lm} \oiint d^2o' X(\mathbf{n},\mathbf{n}') Y_{lm}(\mathbf{n}') =$$
$$= \oiint d^2o' X(\mathbf{n},\mathbf{n}') \sum_{l=0}^{\infty}\sum_{m=-l}^{l} f_{lm} Y_{lm}(\mathbf{n}') = \oiint d^2o' X(\mathbf{n},\mathbf{n}') f(\mathbf{n}').$$

Since $f$ is arbitrary, we conclude that

$$X(\mathbf{n},\mathbf{n}') = \delta^2(\mathbf{n},\mathbf{n}').$$

The spectral expansion (7.4) expresses a well-known fact: the spectrum of an identical operator consists of a single point, unity.

We substitute asymptotic (7.2) into (3.19):

$$\frac{1}{k_0}\frac{\partial u}{\partial r}(R,\mathbf{n}) = \frac{1}{4\pi}\oiint d^2o' u(R,\mathbf{n}') DtN(k_0 R;\mathbf{n},\mathbf{n}'),$$

Singular part (3.24) of the kernel $DtN$ is not localized at large distances because it is a product of a function of $z$-variable and the operator depending on directions $\mathbf{n},\mathbf{n}'$ only. Taking into account the disappearance of the singular part at large distances due to the factor $1/z$, we get:

$$\frac{1}{k_0}\frac{\partial u}{\partial r}(\infty,\mathbf{n}) = \frac{1}{4\pi}\oiint d^2o' u(\infty,\mathbf{n}') K(\infty,\mathbf{n}\cdot\mathbf{n}'),$$

$$\frac{1}{k_0}\frac{\partial u}{\partial r}(\infty,\mathbf{n}) = \frac{1}{4\pi}\oiint d^2o' u(\infty,\mathbf{n}') 2i\delta(1-\mathbf{n}\cdot\mathbf{n}'),$$

$$\frac{1}{k_0}\frac{\partial u}{\partial r}(\infty,\mathbf{n}) = \frac{1}{4\pi}\oiint d^2o' u(\infty,\mathbf{n}') 4\pi i\delta^2(\mathbf{n},\mathbf{n}'),$$

$$\frac{\partial u}{\partial r}(\infty,\mathbf{n}) = ik_0 \oiint d^2o' u(\infty,\mathbf{n}') \delta^2(\mathbf{n},\mathbf{n}'),$$

$$\frac{\partial u}{\partial r}(\infty,\mathbf{n}) = ik_0 u(\infty,\mathbf{n}). \tag{7.5}$$

And this is the Sommerfeld asymptotic (2.2).

**8. Conclusions.** Here, the problem of transferring the Sommerfeld radiation boundary condition from infinity to a sphere of finite radius is solved for a boundary value problem with a three-dimensional Helmholtz equation. The boundary condition has the form of a linear integro-differential relation (3.19), which relates the value of the desired function and its normal derivative on the boundary of the spherical region in a *non-local* way:

$$\frac{1}{k_0}\frac{\partial u}{\partial r}(R,\mathbf{n}) = \frac{1}{4\pi}\oiint d^2o' u(R,\mathbf{n}') K(k_0 R, \mathbf{n}\cdot\mathbf{n}') - \\ -\frac{1}{4\pi k_0 R}\left(\oiint d^2o' u(R,\mathbf{n}') - \oiint d^2o' F(\mathbf{n}\cdot\mathbf{n}') \Lambda' u(R,\mathbf{n}')\right). \quad (8.1)$$

The kernel of the operator on the boundary sphere can be divided into the regular part $K$ (the sum of (3.22) and (3.23), an integral operator) and the singular part $S$ (3.24). The latter, with subtraction the operator of averaging over angles, can be represented as a product of two commuting self-adjoint operators: integral and differential 2nd order (5.14). The regular part of the operator $K$ can be decomposed into the analytical part $A$ (3.25) and the computational part $R$ (3.22), the latter is computable using an absolutely and uniformly convergent series.

The bilinear form (6.1) generated by the singular part (3.24) of the boundary condition operator (3.14) admits a symmetric representation (6.2) containing derivatives of at most 1st order. This favors the future development of the *finite element method* (FEM) for solving boundary value problems with a new type of boundary conditions. This enables to represent the boundary condition in the form (3.19):

$$\frac{1}{k_0}\frac{\partial u}{\partial r}(R,\mathbf{n}) = \frac{1}{4\pi}\oiint d^2o' u(R,\mathbf{n}') DtN(k_0 R; \mathbf{n}, \mathbf{n}'),$$

where the kernel of the Dirichlet-Neumann mapping is defined by the formulas

$$DtN(z;\mathbf{n},\mathbf{n}') = K(z,\mathbf{n}\cdot\mathbf{n}') - \frac{1}{z}\left(1 + \mathrm{div}\left((\nabla\gamma(\mathbf{n}\cdot\mathbf{n}'))b(\mathbf{n}\cdot\mathbf{n}')(\nabla'\gamma(\mathbf{n}\cdot\mathbf{n}'))\right)\cdot\nabla'\right). \quad (8.2)$$

Or in operator form:

$$\mathrm{DtN} = \mathrm{K} - \frac{1}{z}(\mathrm{I} + \mathrm{div}\,\mathrm{M}\,\nabla), \quad (8.3)$$

$$\mathrm{DtN}\,u(R,\mathbf{n}) \equiv \oiint d^2o' u(R,\mathbf{n}') DtN(k_0 R; \mathbf{n}, \mathbf{n}'),$$
$$\mathrm{K}\,u(R,\mathbf{n}) \equiv \oiint d^2o' u(R,\mathbf{n}') K(k_0 R, \mathbf{n}\cdot\mathbf{n}'),$$
$$\mathrm{M}\,\mathbf{v}(\mathbf{n}) \equiv \oiint d^2o' M(\mathbf{n},\mathbf{n}') \mathbf{v}(\mathbf{n}'), \quad (8.4)$$
$$M(\mathbf{n},\mathbf{n}') \equiv (\nabla\gamma(\mathbf{n}\cdot\mathbf{n}')) b(\mathbf{n}\cdot\mathbf{n}') (\nabla'\gamma(\mathbf{n}\cdot\mathbf{n}'))^T.$$

Here the symbols $\mathrm{div}, \gamma, \nabla\gamma, b, \nabla'\gamma$ are defined by formulas (5.10), (3.15), (5.9), (6.18), (5.9) (in primed variables), respectively. The kernel $M$ is a degenerate symmetric real $2\times 2$ matrix. The singular part is reduced to a symmetrical form[8], containing only first-order derivatives $\nabla' u$ of the boundary value of the desired function on the sphere.

This new type of boundary condition is proposed to be called *the boundary condition of the 4th kind*. The boundary condition of the 3rd kind (the Robin condition) is a generalization of the historically initial conditions of the 1st kind (Dirichlet) and the 2nd kind (Neumann). The boundary condition of the 4th kind is a further generalization of the condition of the 3rd kind to the nonlocal case. There is a limit at which the boundary condition of the 4th kind (nonlocal) passes into the boundary condition of the 3rd kind (localized). If the radius $R$ of the boundary sphere has many wavelengths (wave zone), and the radius of the ball $R_1$, in which the radiation source is enclosed, is much less than $R$, then the relationship between the function and its normal derivative is localized, and the nonlocal boundary condition can be approximately replaced by a local condition of the third kind (the Robin condition):

$$\frac{\partial u}{\partial r}(R, \mathbf{n}) \approx i k_0 u(R, \mathbf{n}). \tag{8.5}$$

The latter is known also in the literature as the Sommerfeld radiation condition (2.2).

Currently, such well-known packages as "Wolfram Mathematica" and COMSOL are not prepared for solving problems with non-local boundary conditions.

**9. Acknowledgements.** The author is grateful to his colleagues Dr. Yu. V. Yakovenko and Dr. O. S. Burdo for a useful discussion of the work, as well as Dr.eng. I. S. Petukhov for help in selecting literature.

---

[8] The symmetry is meant in the sense of the generated bilinear form, it is similar to the symmetry of the Legendre operator $(d/d\xi)(1-\xi^2)(d/d\xi)$.